\documentclass[twocolumn,prl,superscriptaddress,showpacs]{revtex4}
\usepackage{amsmath,amssymb}
\usepackage{graphicx}
\usepackage{float}
\usepackage{subfigure}
\usepackage{verbatim}
\usepackage{amsfonts}
\usepackage{dcolumn}
\usepackage{bm}
\usepackage{color}
\usepackage[colorlinks,citecolor=blue]{hyperref}
\usepackage{array}

\begin{document}

\hyphenpenalty=5000
\tolerance=1000

\title{Modular Matrices as Topological Order Parameter\\
by Gauge Symmetry Preserved Tensor Renormalization Approach}

\author{Huan He}
\affiliation{Perimeter Institute for Theoretical Physics, 31 Caroline St N, Waterloo, ON N2L 2Y5, Canada}
\author{Heidar Moradi}
\affiliation{Perimeter Institute for Theoretical Physics, 31 Caroline St N, Waterloo, ON N2L 2Y5, Canada}
\author{Xiao-Gang Wen}
\affiliation{Perimeter Institute for Theoretical Physics, 31 Caroline St N, Waterloo, ON N2L 2Y5, Canada}
\affiliation{Department of Physics, Massachusetts Institute of Technology, Cambridge, Massachusetts 02139, USA}

\begin{abstract}
Topological order has been proposed to go beyond Landau symmetry breaking theory for more than twenty years. But it is still a challenging problem to generally detect it in a generic many-body state. In this paper, we will introduce a systematic numerical method based on tensor network to calculate modular matrices in 2D systems, which can fully identify topological order with gapped edge. Moreover, it is shown numerically that modular matrices, including $S$ and $T$ matrices, are robust characterization to describe phase transitions between topologically ordered states and trivial states, which can work as topological order parameters. This method only requires local information of one ground state in the form of a tensor network, and directly provides the universal data (S and T matrices), without any non-universal contributions. Furthermore it is generalizable to higher dimensions. Unlike calculating topological entanglement entropy by extrapolating, which numerical complexity is exponentially high, this method extracts a much more complete set of topological data (modular matrices) with much lower numerical cost.
\end{abstract}

\maketitle

\section{Introduction}
The most basic question in condensed matter is to classify all different states and phases. Landau symmetry breaking theory is the first successful step to classify all phases  \cite{L3726,L3745,LanL58}. However, the experimental discovery of Integer Quantum Hall Effect \cite{KDP8094} and Fractional Quantum Hall Effect \cite{TSG8259} led condensed matter physics to a new era that goes beyond Landau theory, in which the most fundamental concept is topological order \cite{Wtop,WNtop,Wrig}. Topological order is characterized/defined by a new kind of "topological order parameter": (a) the topology-dependent \emph{ground state degeneracy}  \cite{Wtop,WNtop} and (b) the \emph{non-Abelian geometric phases $S$ and $T$} of the degenerate ground states \cite{Wrig,KW9327,W1221}, where both of them are \emph{robust against any local perturbations} that can break any symmetries \cite{WNtop}. This is just like superfluid order being characterized/defined by zero-viscosity and quantized vorticity that are robust against any local perturbations that preserve the $U(1)$ symmetry.

Recently, it was found that, microscopically, topological order is related to long-range entanglement \cite{LW0605,KP0604}. In fact, we can regard topological order as pattern of long-range entanglement \cite{CGW1038} defined through local unitary (LU) transformations.\cite{LWstrnet,VCL0501,V0705} Chiral spin
liquids, \cite{KL8795,WWZ8913} integral/fractional quantum Hall states \cite{KDP8094,TSG8259,L8395}, $Z_2$ spin liquids, \cite{RS9173,W9164,MS0181} non-Abelian fractional quantum Hall states, \cite{MR9162,W9102,WES8776,RMM0899} are examples of topologically ordered phases.  Topological order and long-range entanglement are truly new phenomena, which require new mathematical language to describe them. It appears that tensor category theory \cite{FNS0428,LWstrnet,CGW1038,GWW1017,GWW1332} and simple current algebra \cite{MR9162,BW9215,WW9455,LWW1024} (or pattern of zeros \cite{WW0808,WW0809,BW0932,SL0604,BKW0608,SY0802,BH0802,BH0802a,BH0882}) may be part of the new  mathematical language.  For 2+1D topological orders (with gapped or gappless edge) that have only Abelian statistics, we find that we can use integer $K$-matrices to classify them.\cite{BW9045,R9002,FK9169,WZ9290,BM0535,KS1193}

As proposed in Ref. \cite{Wrig,KW9327,W1221}, the non-Abelian geometric phases of the degenerate ground states, {\it i.e.} the Modular matrices generated by Dehn twist and 90 degree rotation, are effective "topological order parameters" that can be used to characterize topological order. Refs. \cite{ZGT1251,TZQ1251,ZMP1233,CV1308} makes the first step to calculate numerically modular matrices using various methods. Actually, the relation of tensor network states (TNS) and topological order has already been investigated by several papers \cite{GLS0918,BAV0919}. Ref. \cite{CZG1019,SW1017,SchuchEtal,Buerschaper:2013nga} concluded that gauge-symmetry structure of TNS will give rise to information of topological order. Unlike calculating topological entanglement entropy which in principle needs to calculate the reduced density matrix with exponentially high computational cost, extracting topological data through the gauge-symmetry structure of TNS has acceptable lower cost.

In this paper, we will give a systematical approach to calculate modular matrices, using the wave-function overlap method proposed in Ref. \cite{HW1339,MW13}. Our approach is based on TNS and gauge-symmetry preserved tensor renormalization group. Gauge-symmetry preserved RG differs from original tensor RG (TRG) in the sense that every step of TRG will keep the gauge-symmetry structure invariant and manifest. The paper is organized as follows: I) we will first review the basic ideas of modular matrices and TRG; II) we will explain the systematical method to calculate modular matrices based on TRG; III) we will show the numerical results of modular matrices for the toric code and double-semion topological orders,\cite{RS9173,W9164,MS0181,FNS0428,LWstrnet} which clearly identifies the correct topological order and characterizes phase transitions.

\section{Review of Modular Matrices}
Modular matrices, or $T$- and $S$-matrices, are generated respectively by Dehn twist (twist) and $90^\circ$ rotation on torus. The operation of
twist can be defined by cutting up a torus along one axis, twisting the edge by $360^\circ$ and glueing the two edges back.

The elements of the universal $T$- and $S$-matrices are given by: \cite{HW1339,MW13}
\begin{gather}\label{eq.SE}
\begin{aligned}
\langle\psi_i|\hat T|\psi_j\rangle
&=& e^{-A/\xi^2+o(1/A)}
  T_{ij}
\\
\langle\psi_i|\hat S|\psi_j\rangle
&=& e^{-A/\xi^2+o(1/A)}
  S_{ij}
\end{aligned}
\end{gather}
where $|\psi_i\rangle$ form a set of orthonormal basis for degenerate ground space; and $\hat T$ and $\hat S$ are the operators that generate the twist
and the rotation on torus. $A$ is the area of the system and $\xi$ is of order of correlation length which is not universal.

The $T$- and $S$-matrices encode all the information of quasi-particles statistics and their fusion.\cite{Wang10,LW1384} It was also conjectured that the $T$- and $S$-matrices form a complete and one-to-one characterization of topological orders with gapped edge \cite{Wrig,KW9327,W1221} and can replace the fixed-point tensor description to give us a more physical label for topological order.

\section{Review of Tensor Renormalization Group}
To be specific, TRG here means double tensor renormalization group \cite{GLW0816}. Essentially, a translation invariant TNS can be written by definition as
\begin{equation}
|\psi\rangle=\sum_{m_1m_2..}\text{tTr}(T^{m_1}T^{m_2}...T^{m_N})|m_1\rangle|m_2\rangle...|m_N\rangle
\end{equation}
where $T^{m_i}$'s are local tensors with physical index $m_i$ defined either on links or vertices; and $m_i$'s are local Hilbert space basis. (Sometimes $m_i$ is not written out explicitly if there is no ambiguity). $\text{tTr}$ means contracting over all internal indices of local tensors pair by pair. The norm of the state is given by
\begin{equation}
    \langle\psi|\psi\rangle=\text{tTr}(\boldsymbol{T}\boldsymbol{T}...\boldsymbol{T})
\end{equation}
where, $\boldsymbol{T}$ is the local double tensor, which is formed by $T^\star$ and $T$ tracing out physical degree freedom.
\begin{equation}
    \boldsymbol{T}=\sum_{m_i} T^{m_i\star}T^{m_i}
\end{equation}

The essence of double TRG is to find fewer double tensors $\boldsymbol{T}^\prime$, which keeps the norm approximately invariant. I.e.,
\begin{equation}
    \langle\psi|\psi\rangle\simeq \text{tTr}(\boldsymbol{T}^\prime\boldsymbol{T}^\prime...\boldsymbol{T}^\prime)
\end{equation}

\begin{figure}
  \includegraphics[width=0.5\textwidth]{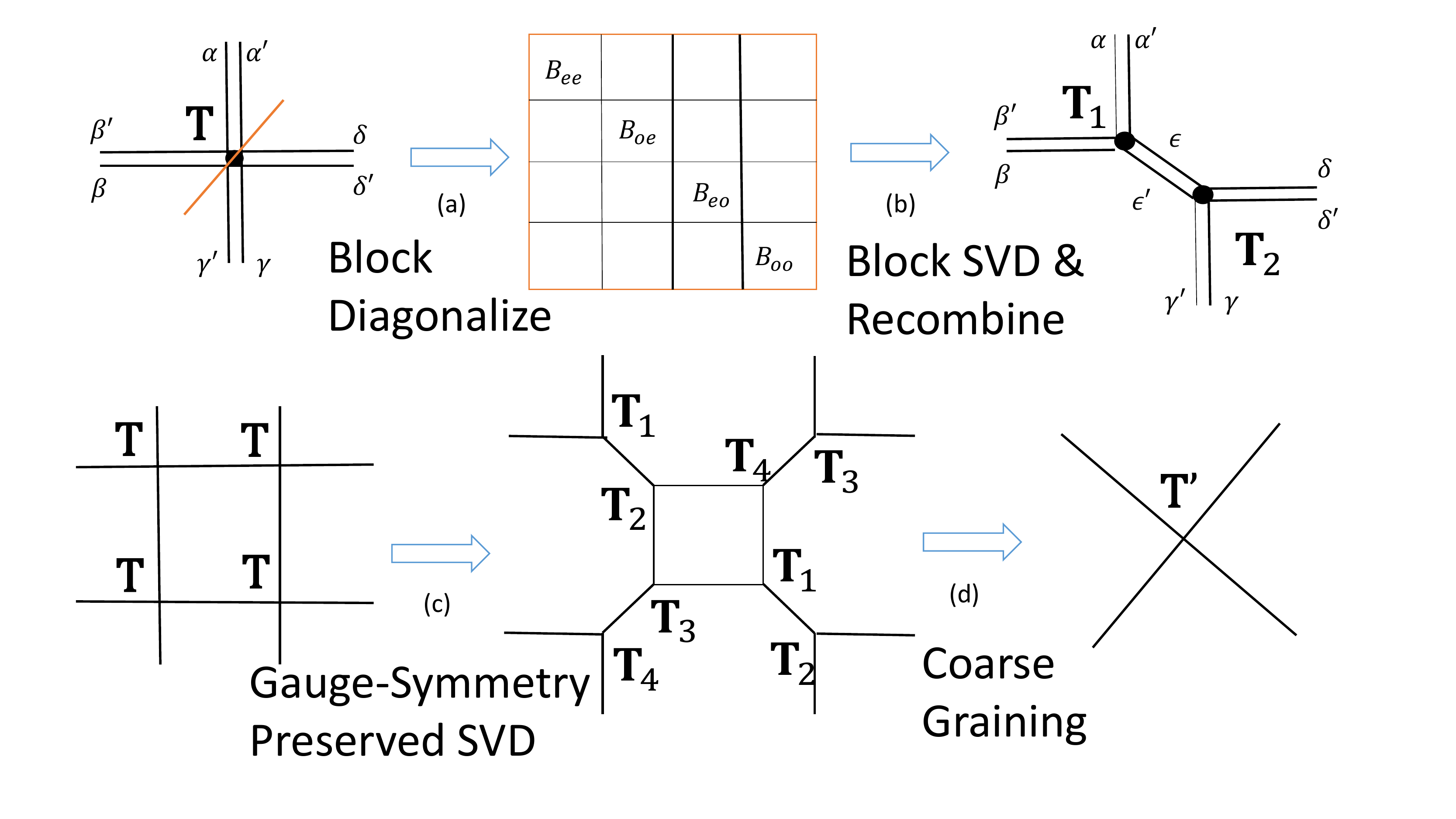}\\
  \caption{Illustration for Symmetry Preserved Tensor Renormalization Group. First (a) before SVD, block diagonalize double tensor $\boldsymbol{T}$ according to the $Z_2$ symmetry rule, $\alpha+\beta+\gamma+\delta$ and $\alpha^\prime+\beta^\prime+\gamma^\prime+\delta^\prime$ are both even numbers. Therefore the indices of each block matrices $B_{ee}$, $B_{eo}$, $B_{oe}$, $B_{oo}$ represent whether $\alpha+\beta$ and $\alpha^\prime+\beta^\prime$ are even or odd. (b) Perform SVD in each block matrices and recombine the tensors coming out of SVD into tensor $\boldsymbol{T}_1$ and $\boldsymbol{T}_2$, according to the rule $\alpha+\beta+\epsilon^\prime$, $\alpha^\prime+\beta^\prime+\epsilon$, $\gamma+\delta+\epsilon^\prime$ and $\gamma^\prime+\delta^\prime+\epsilon$ are all even numbers. I.e., tensor $\boldsymbol{T}_1$ and $\boldsymbol{T}_2$ both obey $Z_2$ gauge symmetry. (c) and (d) are the same procedures as TRG. (c) is to use SVD to decompose $\boldsymbol{T}$ into $\boldsymbol{T}_1$ and $\boldsymbol{T}_2$. Only $D_{cut}$ numbers of singular values will be kept. (d) is coarse graining. The four tensors on the small square will form a new double tensor $\boldsymbol{T}^\prime$. Note that $T_3$ and $T_4$ are outcoming tensors that are cut in another direction.}
  \label{fig.RG}
\end{figure}

This approximation can be done non-uniquely. And SVD TRG approach shall be utilized in this paper for its convenience and low cost. The procedure of SVD RG approach is graphically explained in the Fig. \ref{fig.RG} (c) and (d). Step (c) is to perform local SVD to decompose double tensor $\boldsymbol{T}$ into $\boldsymbol{T}_1$ and $\boldsymbol{T}_2$. In order to prevent the bond dimension of internal indices from growing exponentially, only finite number $D_{cut}$ of singular values are kept. Step (d) is to do coarse graining, the tensors on new smaller square will form a new double tensor $\boldsymbol{T}^\prime$. After step (c) and (d), half of tensors will be contracted. For a translation invariant TNS, after enough steps of SVD TRG, the double tensor will flow to the fixed point double tensor, $\boldsymbol{T}_{fp}$, which plays an essential role in the next section. Topological data can be extracted from $\boldsymbol{T}_{fp}$.

Note that the above TRG approach suffers from the necessary symmetry condition \cite{CZG1019}. If the gauge symmetry is not preserved in each step of TRG, the approach will be ruined by errors. And more importantly, the RG flow will arrive at some wrong fixed point tensors. Gauge-Symmetry Preserved TRG is introduced in the next section in order to prevent this happening. Another reason that normal TRG is not suitable here is that during TRG, the gauge symmetry information is lost. So that in order to reproduce all topological data, the gauge symmetry should be preserved.

\section{Modular Matrices by Gauge-Symmetry Preserved Tensor Renormalization Group}
In refs. \cite{SW1017,SchuchEtal,Buerschaper:2013nga}, the gauge structure of TNS is analyzed. It was concluded that by inserting gauge transformation tensors to TNS, a set of basis for the degenerate ground space will be obtained. More specifically, the ground states could be labeled as $|\psi(g,h)\rangle$, where $g,h$ are gauge tensors acting on internal indices in two directions. Different ground states can be transformed to each other by applying gauge tensors on internal indices of a TNS. Therefore it is natural to think that since all ground states could be obtained, by calculating all overlaps $\langle\psi_i|\hat T|\psi_j\rangle$ and $\langle\psi_i|\hat S|\psi_j\rangle$, the whole modular matrices could be calculated. However, it is difficult to compute the overlap directly and keep track of the non-universal contributions. See EQ. \ref{eq.SE}.

TRG will help reduce the difficulty, since one fixed point double tensor essentially represents the whole lattice. Calculating on one double tensor is much easier and size effects do not appear. However, normal TRG is not suitable here since gauge symmetry needs to be preserved through every tensor RG step in order to insert gauge transformation tensors.

To be more specific, let us consider the case of $Z_2$ topological order, which also makes it clear in the next section. As already known in the refs. \cite{SW1017,SchuchEtal,Buerschaper:2013nga}, tensor network representation for $Z_2$ topological state has $Z_2$ gauge symmetry.  The double tensor $\boldsymbol{T}_{\alpha\alpha^\star\beta\beta^\star\gamma\gamma^\star\delta\delta^\star}$ will have a $Z_2\times Z_2$ gauge symmetry, where
$\alpha,\alpha^\star,\beta,\beta^\star,\gamma,\gamma^\star,\delta,\delta^\star = 0,1$, and $\alpha,\beta,\gamma,\delta$ are indices coming from $T$ while $\alpha^\star,\beta^\star,\gamma^\star,\delta^\star$ are indices coming from $T^\star$. So the double tensor with $Z_2$ gauge symmetry satisfies
\begin{equation}
\begin{split}
&   \boldsymbol{T}_{\alpha^\prime\alpha^{\star\prime}\beta^\prime\beta^{\star\prime}\gamma^\prime\gamma^{\star\prime}\delta^\prime\delta^{\star\prime}}=
    \boldsymbol{T}_{\alpha\alpha^\star\beta\beta^\star\gamma\gamma^\star\delta\delta^\star} \times  \\
&   A_{\alpha\alpha^\prime}A_{\beta\beta^\prime}A_{\gamma\gamma^\prime}
    A_{\delta\delta^\prime}B_{\alpha^\star\alpha^{\star\prime}}B_{\beta^\star\beta^{\star\prime}}B_{\gamma^\star\gamma^{\star\prime}}
    B_{\delta^\star\delta^{\star\prime}}
\end{split}
\end{equation}
where repeated indices imply summation and tensor $A,~B~\in \{I,~\sigma_z\}$ generate the $Z_2\times Z_2$ gauge symmetry on both layer of double tensor, which only act on internal indices. If a double tensor has such a gauge symmetry, its elements are nonzero only when $\alpha+\beta+\gamma+\delta$ and
$\alpha^\star+\beta^\star+\gamma^\star+\delta^\star$ are both even. \footnote{For general $Z_N$ model, the generators are $\{(A)_{\alpha\beta}=e^{i\frac{2\pi}{N}c\beta}\delta_{\alpha\beta}\}^{N-1}_{c=0}$. And due to $Z_N$ gauge symmetry, the tensor will satisfy that only the components which summation of indices equal to 0 (mod N) will be nonzero.}

In order to keep $Z_2\times Z_2$ gauge symmetry manifest at each RG step, we develop \textit{gauge-symmetry preserved tensor RG} (GSPTRG). Essentially, it differs from normal TRG only when we do SVD. The double tensor needs to be block diagonalized by even or odd of its indices, and then SVD is performed in each block and recombine the tensors coming out of SVD into one tensor, just as the way to block diagonalize it. In each block, the tensor elements have the same even or odd indices, which therefore is key to preserve $Z_2$ symmetry manifest. The procedures are also explained in the Fig. \ref{fig.RG}.

\begin{figure}
  \includegraphics[width=0.5\textwidth]{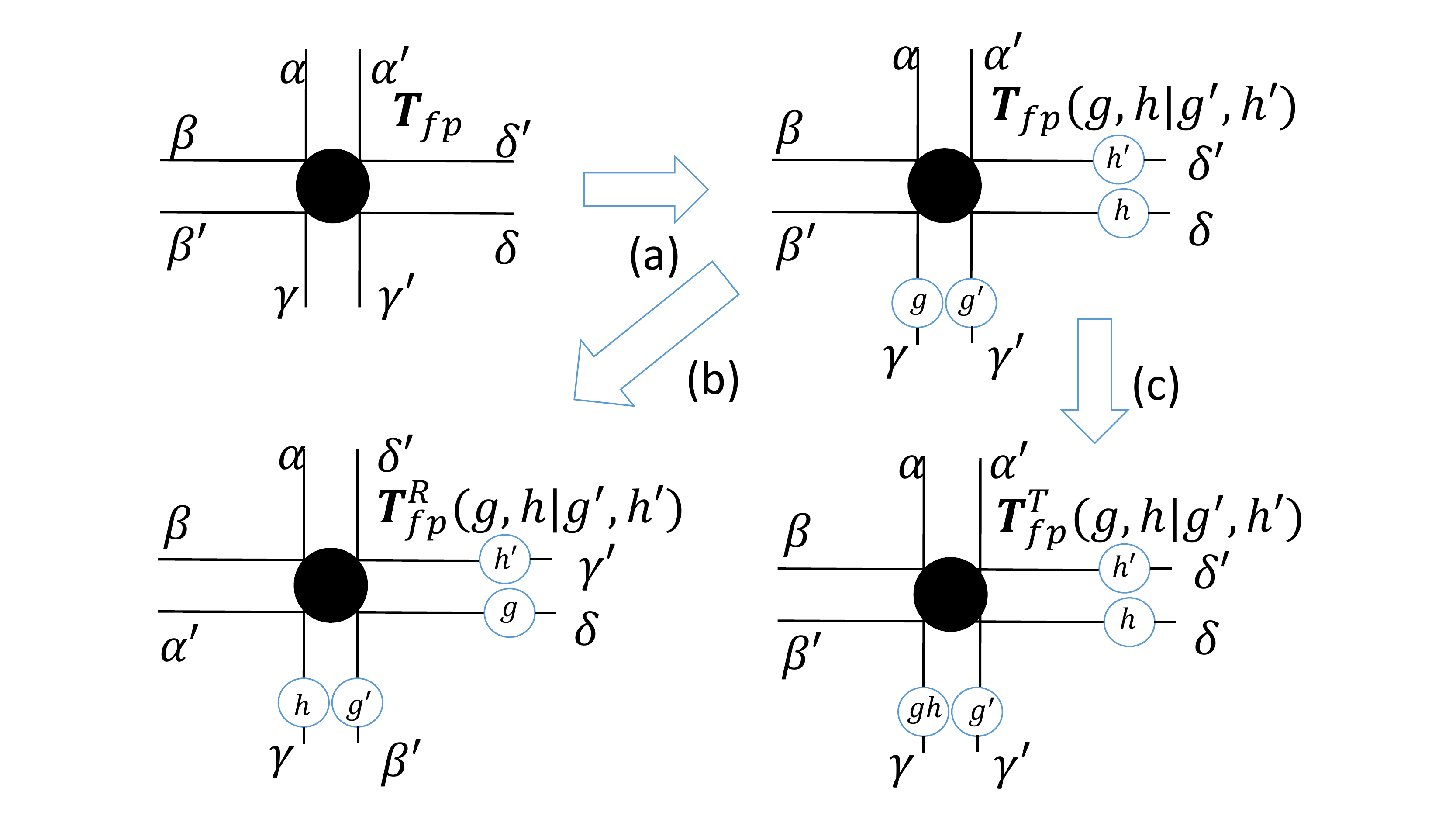}\\
  \caption{Modular matrices from the fixed point double tensor $\boldsymbol{T}_{fp}$. Eight legs of $\boldsymbol{T}_{fp}$ will all be traced over because of torus geometry. (a) By inserting $Z_2$ gauge tensors $g,h,g^\prime,h^\prime$ into $\boldsymbol{T}_{fp}$, $\boldsymbol{T}_{fp}(g,h|g^\prime,h^\prime)$ is obtained; and tracing over eight legs of $\boldsymbol{T}_{fp}(g,h|g^\prime,h^\prime)$ will give rise to overlaps of $\langle\psi(g^\prime,h^\prime)|\psi(g,h)\rangle$, where $|\psi(g,h)\rangle$ labels different ground states with gauge symmetry on boundary. The elements of $T$- and $S$-matrices are just reshuffling of $\langle\psi(g^\prime,h^\prime)|\psi(g,h)\rangle$, as illustrated in the Fig. (b) and (c). Fig. (b) represents $90^\circ$ rotation and Fig. (c) represents twist.}
  \label{fig.bc}
\end{figure}

After several steps of GSPTRG (c.f. Fig. \ref{fig.PD}), double tensor will flow to the gauge-symmetry preserved fixed point tensor. Equivalent to calculate the overlap by brute force, we can obtain the modular matrices by the following three steps:

1) inserting gauge symmetry tensors into double tensor;
2) performing rotation and twist on one layer of fixed point double tensor;
3) tracing out rest indices.

The procedures are also explained in the Fig. \ref{fig.bc}. Actually the innerproduct of ground states $(\langle\psi(g^\prime,h^\prime)|\psi(g,h)\rangle)$ (each ground state is obtained by inserting gauge tensors on boundary) in topological phase will be diagonal with each element modulo 1. The elements of $T$- and $S$-matrices are just reshuffle of elements $(\langle\psi(g^\prime,h^\prime)|\psi(g,h)\rangle)$. More explicitly for the $Z_2$ topological state
\begin{eqnarray}\label{eq.shuffle}
  \langle\psi(g^\prime,h^\prime)|\hat T|\psi(g,h)\rangle &=& \langle\psi(g^\prime,h^\prime)|\psi(g,gh)\rangle\\
  \langle\psi(g^\prime,h^\prime)|\hat S|\psi(g,h)\rangle &=& \langle\psi(g^\prime,h^\prime)|\psi(h,g^{-1})\rangle
\end{eqnarray}

\section{Modular Matrices for $Z_2$ topological order}
Toric code model\cite{K032} is the simplest model that realize the $Z_2$ topological order.\cite{RS9173,W9164} Local physical states are defined on every link with spin up and down. In the notation of string-net states, spin up represents a string while spin down represents no-string. Essentially, the $Z_2$ topological state can be written as equal superposition of all closed string loops:
\begin{equation}
    |\psi_{TC}\rangle=\sum_X |X\rangle
\end{equation}
where X represents a closed loop, and normalization factor is implicity in the above equation.

\begin{figure}
  \includegraphics[width=0.55\textwidth]{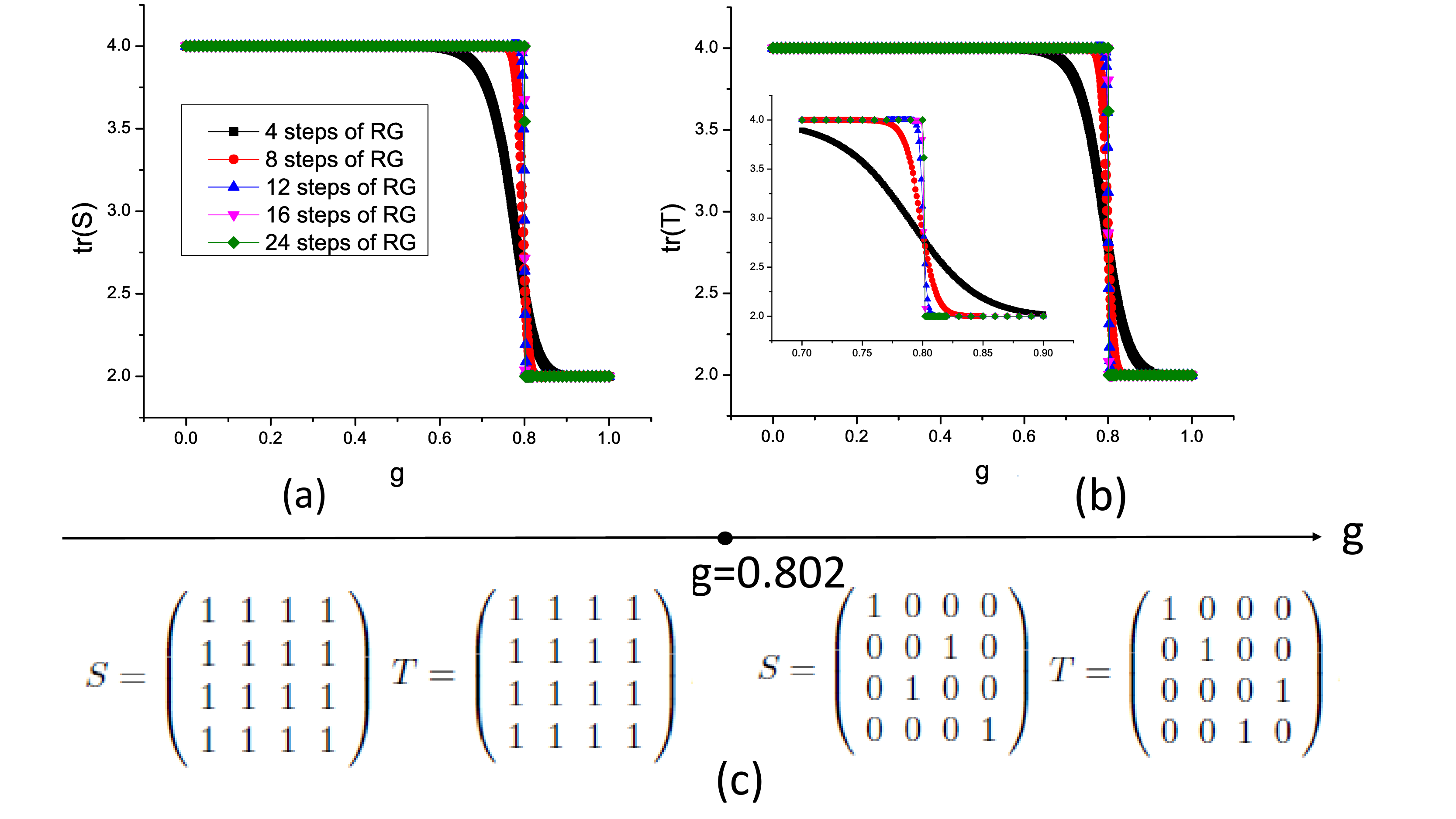}\\
  \includegraphics[width=0.55\textwidth]{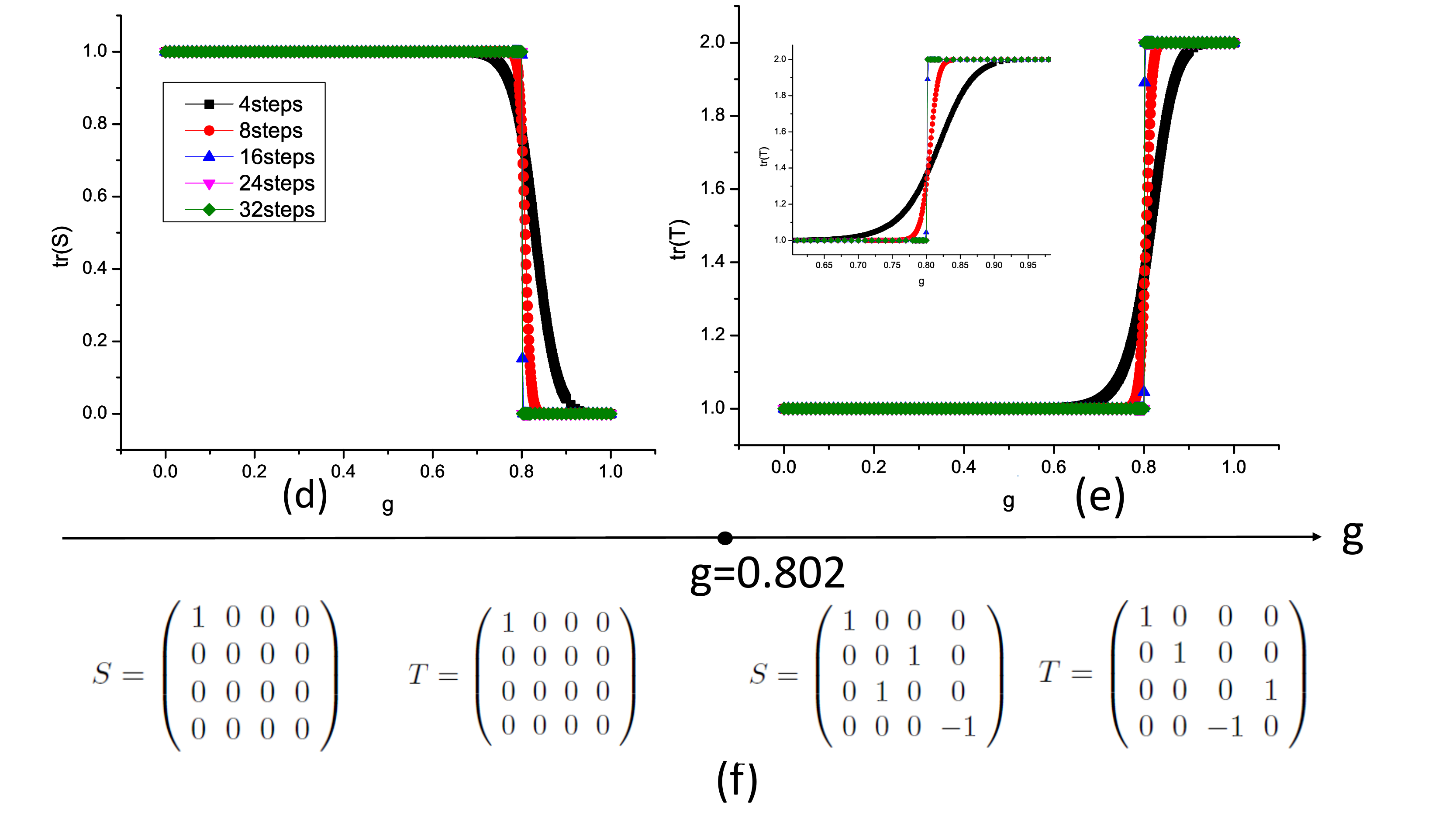}\\
  \caption{
The trace of modular matrices $S$ and $T$ as functions of $g$ display a very
sharp phase transition at critical point $g_c$ as increasing RG steps, for both
$Z_2$ and double-semion topological order. The $Z_2$ topological order
transition point coincides exactly with the results in Ref. \cite{CGW1038} by
another characterization.}
  \label{fig.PD}
\end{figure}

When putting the $Z_2$ topologically ordered state on a torus, the ground state degeneracy is four and the quasi-particles are usually labeled by $\{1,e,m,em\}$. $T$- and $S$-matrices in the twist basis\cite{MW13} are given in Fig. \ref{fig.PD}c for $g>0.802$.

It is easy to represent $|\psi_{TC}\rangle$ in terms of a tensor network.  For the sake of convenience, we replace local physical states $|1\rangle$ and
$|0\rangle$ with $|11\rangle$ and $|00\rangle$ respectively. And combine each $|1\rangle$ and $|0\rangle$ to its nearest sites. So local physical states now
are on vertices without extending Hilbert space. Here we choose the parameterization of $Z_2$ topological state utilized in Ref. \cite{CGW1038}
\begin{equation}\label{eq:MM}
\begin{split}
& T^{(\alpha\beta\gamma\delta)}_{\alpha\beta\gamma\delta}=g^{\alpha+\beta+\gamma+\delta}
\text{when}~~\alpha+\beta+\gamma+\delta~~\text{even} \\
\notag
& \text{Rest~~elements~~of~~T~~are~~zeros}.
\end{split}
\end{equation}
When $g=1$, it is $|\psi_{TC}\rangle$ while when g=0, it is a trivial state $|0000....0\rangle$. Of course, when $g$ is driven from 0 to 1, it must undergo a phase transition.

We calculate $T$- and $S$-matrices along $g$. We find that when $0\leq g<0.802$, all components of $T$- and $S$-matrices are 1, because the gauge twisting does not produce other ground states in the trivial phase. When $0.802\leq g<1$, it belongs to $Z_2$ topological phase, since the $T$- and $S$-matrices for each $g\in (0.802,1]$ agrees with that of $Z_2$ topological phase\cite{MW13} (see Fig. \ref{fig.PD}c).

\begin{figure}
  \includegraphics[width=0.45\textwidth]{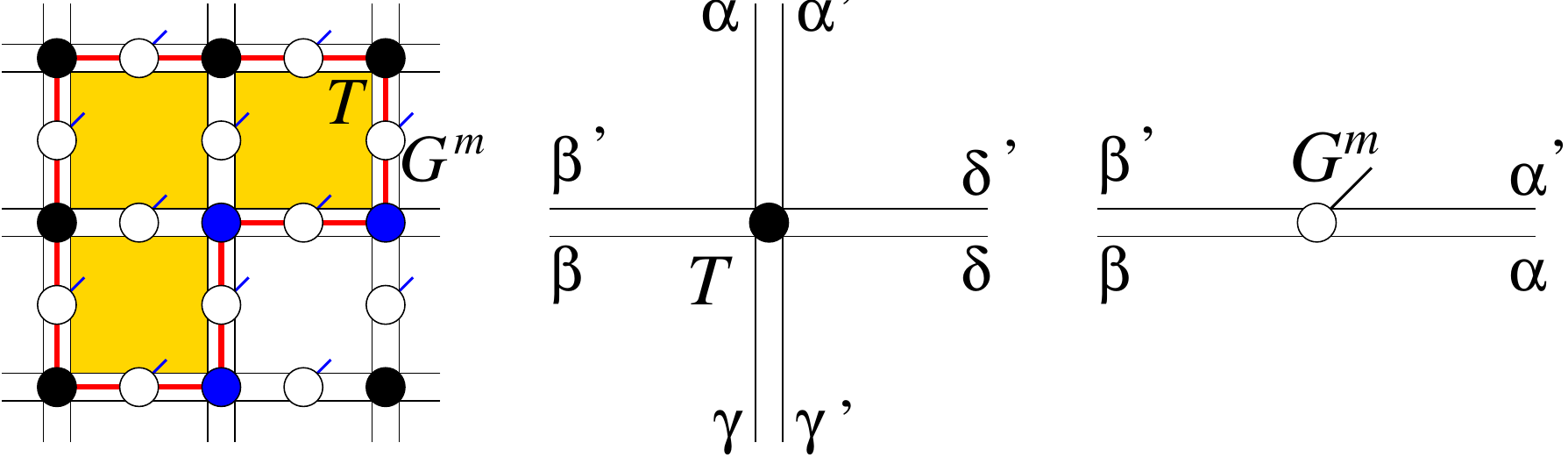}\\
  \caption{The $T$ tensor and the $G^m$ tensor that describes the ground state
wave function of the double semion model.  The ``virtual qubits'' are in the
``1'' state in the shaded squares and in the ``0'' state in other squares.  The
red line is the domain wall (string) between ``0'' and ``1'' states of the
virtual qubits.  The blue (black) dots represent $t_{\alpha\beta\gamma'\delta'}=-1$ ($t_{\alpha\beta\gamma'\delta'}=1$).
}
  \label{dsTG}
\end{figure}

\section{Modular Matrices for Double-semion model}
The double-semion model\cite{FNS0428,LWstrnet,YW} is another topologically ordered state with two semions of statistics $\theta =\pm \pi/2$.  In the notation of string-net states, the double-semion ground state can also be written as superposition of all closed string loops:
\begin{equation}
    |\psi_{DS}\rangle=\sum_X (-)^{N_{loops}}|X\rangle
\end{equation}
where X represents a closed loop, and $N_{loops}$ the number of loops. The above double-semion state can be described by a TNS with the following tensors $T$ and $G^m$ at $g=1$ (see Fig. \ref{dsTG}):
\begin{align}
 T_{ (\alpha \alpha') (\beta \beta') (\gamma \gamma') (\delta \delta')}
&= t_{\alpha\beta\gamma'\delta'}
\delta_{\alpha\beta'}
\delta_{\beta\gamma}
\delta_{\gamma'\delta}
\delta_{\delta'\alpha'} ,
\nonumber\\
 t_{1000} = t_{1101}&=-1,\ \
\text{other } t_{\alpha\beta\gamma'\delta'} =1;
\nonumber\\
G^m_{(\alpha \alpha') (\beta \beta')}&=g^m_{\alpha\alpha'}
\delta_{\alpha\beta}
\delta_{\alpha'\beta'} ,
\nonumber\\
g^1_{10}&= g^1_{01}=g,\ \ \ g^0_{00}= g^0_{11}=1,
\nonumber\\
g^1_{00}&= g^1_{00}= g^0_{10}= g^0_{01}=0.
\end{align}
Note that if we view $\alpha=\beta'$, $\beta=\gamma$, $\gamma'=\delta$, and $\delta'=\alpha'$ as indices that label ``virtual qubits'' in the squares, then
the strings can be viewed as domain wall between the "0" and "1" states of the qubits.  Also if we choose $t_{\alpha\beta\gamma'\delta'} =1$, the above
tensors will describe the $Z_2$ topologically ordered state discussed previously.

The $Z_2$ gauge symmetry is generated by $\sigma^x\otimes \sigma^x$ acting on each internal indices $(\alpha\alpha')$ followed by a transformation generated
by $u^i_{\alpha\alpha'}$, $i=t,l,b,r$ acting on the links of the four orientations. Here $u^i_{\alpha\alpha'}$ must satisfy
\begin{align}
 f_{\alpha\beta\gamma'\delta'}=
u^t_{\beta\gamma'} u^b_{\alpha\delta'}
u^l_{\beta\alpha}u^r_{\gamma'\delta'}
\end{align}
where
\begin{align}
 f_{1000} = f_{0111} = f_{0010} = f_{1101} &=
-1,\ \
\text{others } f_{\alpha\beta\gamma'\delta'}  =1.
\end{align}
Furthermore $u^i_{\alpha\alpha'}$ must also satisfy
\begin{align}
g^m_{\alpha\alpha'} &= (u^t_{\alpha\alpha'})^* g^m_{\alpha\alpha'} (u^b_{\alpha\alpha'})^*
\nonumber\\
g^m_{\alpha\alpha'} &= (u^l_{\alpha\alpha'})^* g^m_{\alpha\alpha'} (u^r_{\alpha\alpha'})^* .
\end{align}
We find that
\begin{align}
 u^t=u^b=
 \left(
\begin{array}{cc}
1 & -1 \\
1 & 1 \\
\end{array}
\right), \ \ \
 u^r=u^l=
 \left(
\begin{array}{cc}
1 & 1 \\
-1 & 1 \\
\end{array}
\right), \ \ \
\end{align}
See \cite{Buerschaper:2013nga} for a general analysis of twisted gauge structures.

After the GSPTRG calculation, we find a phase transition at $g_c=0.802$. The $S$- and $T$-matrices for the nontrivial phase with $g\in (0.802,1]$ are given by Fig. \ref{fig.PD}f, which agree with the modular matrices for the double semion model in string basis \cite{LWY1329}. For the trivial phase near $g=0$, the modular matrices become
$T_{\alpha\beta}=S_{\alpha\beta}=\delta_{\alpha,0}\delta_{\beta,0}$.

\section{Conclusion}
We have developed a systematic approach, \textit{gauge-symmetry preserved tensor renormalization}, to calculate modular matrices from a generic many-body wave function described by a tensor network.  The modular matrices can be viewed as very robust "topological order parameters" that only change at phase transitions.  The tensor network approach gives rise to $S$ and $T$ matrices in a particular basis which is different from the standard quasipartical basis.\cite{Wrig,KW9327,Wang10,W1221,ZGT1251,TZQ1251,ZMP1233,CV1308,LW1384} The trivial phase will result in trivial modular matrices $S=1$ and $T=1$
(since there is no degeneracy on a torus), and the topological phase will give rise to nontrivial modular matrices, which contain topological informations, such as quasi-particles information, like statistic angle, fusion rule, quantum dimension, etc.

In particular, a general algorithm can be developed: the tensor network ansatz can be imposed with gauge symmetry $G$ (or MPO symmetry, see below) in the beginning, and the corresponding update algorithm, which is used to find ground states, also preserves such a gauge symmetry. Therefore if the topological phase indeed has such a gauge theory description, the ansatz obviously is better than the normal tensor network ansatz. In Appendix B we perform such a benchmark computation using the $Z_2$ phase of the Kitaev honeycomb model \cite{K062}. There we prepare an arbitrary tensor with $Z_2$ symmetry, find the ground state (locally) numerically by gauge-symmetry preserved update and from there compute the modular matrices. A similar tensor network computation of Kitaev honeycomb model is developed in Ref. \cite{HuanHe} where $Z_2$ gauge structure is also imposed but expressed by Grassmann tensor network. The energy and nearest neighbor correlation are computed there.

After the completion and publication of the preprint of this paper, the notion of (twisted) $G$-injectivity of \cite{SchuchEtal,Buerschaper:2013nga} was generalized to the matrix product operator (MPO) case in \cite{MPOgeneralization} and it was shown that any string-net model is included with this generalization. The method developed in this paper can thus similarly be generalized to any MPO symmetry and does not need any group structure (and thus not restricted to twisted discrete gauge theories).

The universal wave function overlap \cite{MW13} \eqref{eq.SE} applies to any dimension and have already been investigated in exactly solvable models in 3+1D \cite{JiangEtal3D,MoradiWen3D,WangWen3D}. The method outlined in this paper can similarly be generalized to higher dimensions to extract universal topological information from generic gapped ground states.

Finally we note that although the universal wave function overlap \cite{MW13} works for any topological order, the machinery developed in this paper in 2+1D only works for non-chiral topological order (gapped boundaries) as formulated here. This is only because the tensor network techniques used are best understood for non-chiral topological order, but a generalization for chiral topological order would be both interesting and important.

The authors appreciate helpful discussions with Lukasz Cincio, Guifre Vidal, Zheng-Cheng Gu, Tian Lan, Fang-Zhou Liu and Oliver Buerschaper.  This research is supported by NSF Grant No.  DMR-1005541, NSFC 11074140, and NSFC 11274192. It is also supported by the John Templeton Foundation.  Research at Perimeter Institute is supported by the Government of Canada through Industry Canada and by the Province of Ontario through the Ministry of Research.

\bibliography{./local,./wencross,./all,./publst}

\clearpage
\appendix
\setcounter{figure}{0}
\renewcommand\thefigure{A.\arabic{figure}}
\setcounter{equation}{0}
\renewcommand\theequation{A.\arabic{equation}}

\section{Appendix A: Robustness of modular matrices under $Z_2$ perturbations}

In the phase diagram Fig. \ref{fig.PD}, it already demonstrates that $T$- and $S$-matrices are very robust characterization of topological order, which only depend on the phase. In order to address on this issue more explicitly, we will perturb $Z_2$ topological state at $g=1$, while the perturbation also respects internal $Z_2$ gauge symmetry, i.e., the perturbation tensor $T^\prime$ is written as:
\begin{equation}
\begin{split}
& T^{\prime\alpha^\prime\beta^\prime\gamma^\prime\delta^\prime}_{\alpha\beta\gamma\delta}=\epsilon r~~when~~\alpha+\beta+\gamma+\delta~~even\\
\end{split}
\end{equation}
where r is a uniform distributed random number ranging from $[-1,1]$ depending on $\alpha^\prime,\beta^\prime,\gamma^\prime,\delta^\prime,\alpha,\beta,\gamma,\delta$; and $\epsilon$ represents perturbation strength starting from zero. The initial tensor before RG will be $T+T^\prime$.

\begin{figure}
  \includegraphics[width=0.4\textwidth]{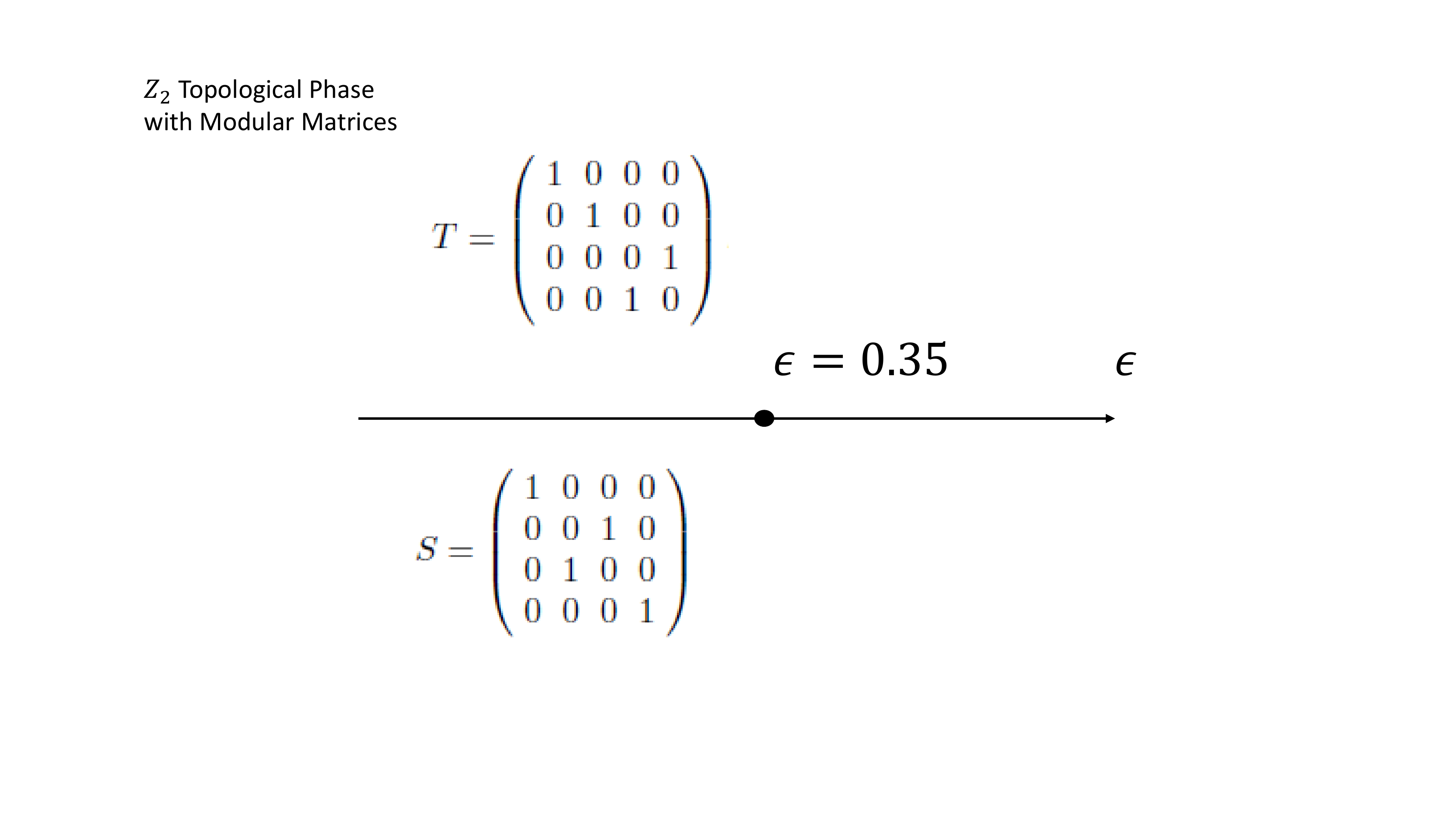}\\
  \caption{Phase diagram under perturbation}
  \label{fig.perturb}
\end{figure}

As already shown in Ref. \cite{CGW1038} paper, $Z_2$ topological phase is robust under tensor perturbations which respect the $Z_2$ gauge symmetry, while fragile under perturbations breaking the $Z_2$ gauge symmetry. Here we start from perturbed tensor $T+T^\prime$ and calculate modular matrices for different $\epsilon$'s, which will demonstrate the robustness of this topological
characterization\ref{fig.perturb}.

Numerically it demonstrates that when $0\leq\epsilon\leq0.35$, $T$- and $S$-matrices are always eqn. \ref{eq:MM}. However, when $\epsilon>0.35$, the perturbations will possibly break the topological phase (and possibly not). In this case, $T$- and $S$-matrices have three possibilities as shown in the figure. Anyway, this calculation clearly demonstrates modular matrices are robust characterization of topological phase.

\appendix
\setcounter{figure}{0}
\renewcommand\thefigure{B.\arabic{figure}}
\setcounter{equation}{0}
\renewcommand\theequation{B.\arabic{equation}}

\section{Appendix B: Gauge-Symmetry Preserved Update}

For a typical tensor network algorithm, there are two main steps: updating local tensors to lower the energy to ground state energy and contracting all local tensors to compute physical quantities and norms. Here we only point out some details in gauge-symmetry preserved update algorithm, since the details in contraction have already been reviewed in the main text to some extent.

We choose Kitaev honeycomb model as a benchmark. Kitaev honeycomb model is defined on the honeycomb lattice with spins on each site and different interactions along the three different links connected to each site
\[H=-J_x\sum_{x-\text{links}} \sigma_i^x\sigma_j^x-J_y\sum_{y-\text{links}} \sigma_i^y\sigma_j^y-J_z\sum_{z-\text{links}} \sigma_i^z \sigma_j^z.\]
%\[H=-\sum_{\langle i j \rangle_\gamma} J_\gamma \sigma_i^\gamma \sigma_j^\gamma,\] with $\gamma=x,y,z$.
$J_\gamma$ are coupling constants along the $\gamma-$link. For simplicity we will assume they are all positive. For the coupling constants $J_\gamma$ satisfying $J_x+J_y<J_z$ (or other permutations), a gapped phase will be acquired that indeed is a toric code phase by perturbation analysis \cite{K062}.

\begin{figure}[H]
\centering
\includegraphics[width=0.4\textwidth]{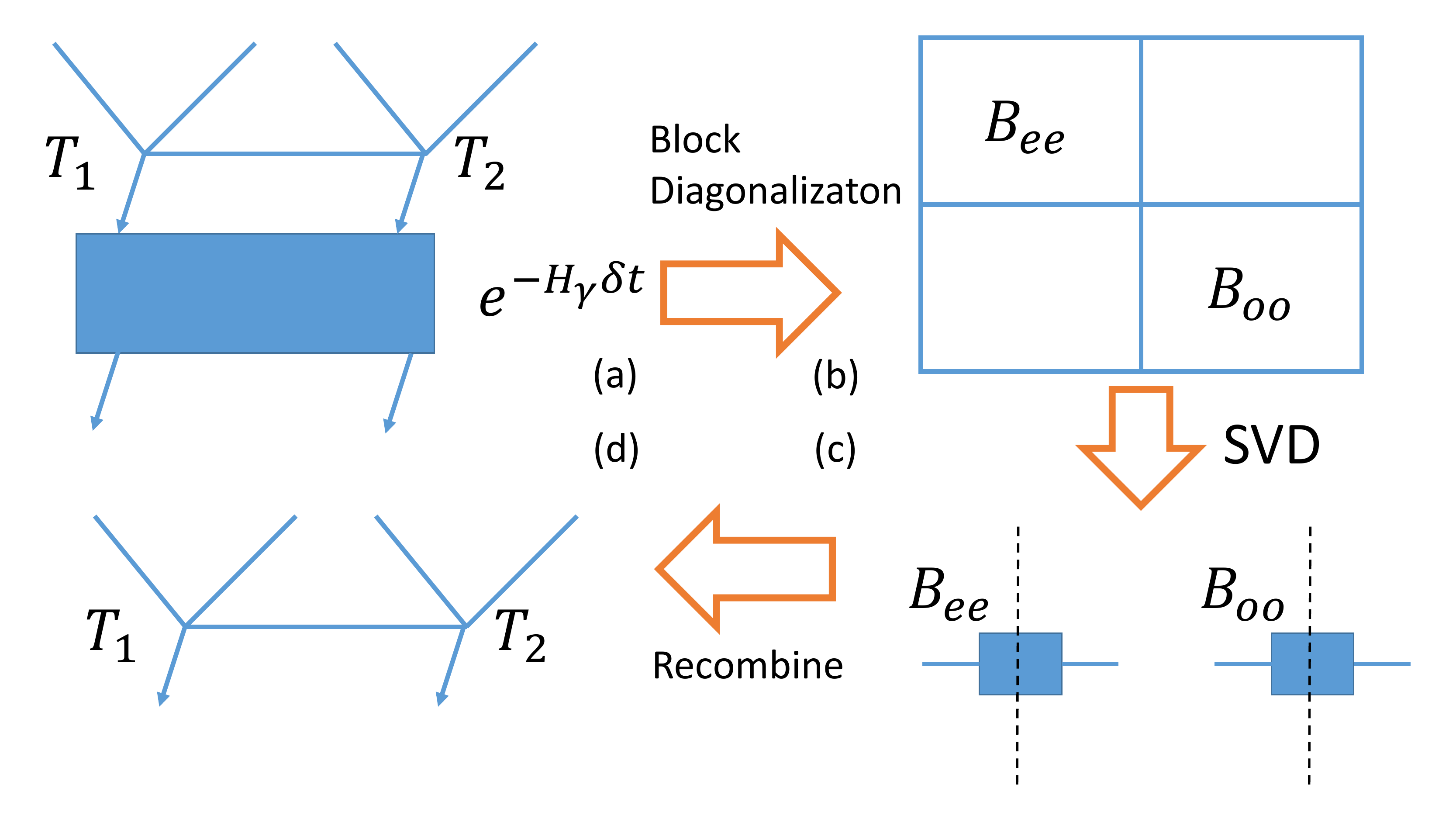}
\caption{Illustration of gauge-symmetry preserved simple update. (a) shows that tenor $T_1$ and $T_2$ are contracted and acted with local imaginary evolution operator represented by the blue box. The legs with arrow are physical indices while legs without arrows are internal indices. (b) Block diagonalization according to internal indices. $B_{ee}$ and $B_{oo}$ represent the matrices with both legs even and odd. (c) $B_{ee}$ and $B_{oo}$ are SVD-ed. (d) The outcoming matrices are recombined into the original form as in figure (a).}
\label{fig.simpleupdate}
\end{figure}

We impose $Z_2$ gauge symmetry on our tensor network ansatz. I.e., local tensors should satisfy
\begin{equation}\label{eq.Z2ansatz}
T^m_{ijk}=0,\quad\text{if}\quad i+j+k\;\text{odd}.
\end{equation}
Other elements of tensors are random in the initial states before simple update. Gauge-symmetry preserved update differs from simple update only when we do SVD. Again, what we need to do in SVD approach is the following three steps: block diagonalization according to gauge symmetry, SVD in each block and rearrange the outcoming tensors back to the original form. Note that the gauge symmetry only acts on internal indices, so that block diagonalization only happens for internal indices. The procedure is also summarized in the figure \ref{fig.simpleupdate}.

We randomly pick up a few points in the gapped phase of Kitaev honeycomb model, use gauge-symmetry preserved update to obtain the ground states by $Z_2$ symmetric ansatz \ref{eq.Z2ansatz}. Modular matrices are calculated by the method explained in the main text, and the result is exactly the same matrices found in the main text:
\begin{eqnarray}
S=
\left(
  \begin{array}{cccc}
    1 & 0 & 0 & 0 \\
    0 & 0 & 1 & 0 \\
    0 & 1 & 0 & 0 \\
    0 & 0 & 0 & 1
  \end{array}
\right),
T=
\left(
  \begin{array}{cccc}
    1 & 0 & 0 & 0 \\
    0 & 1 & 0 & 0 \\
    0 & 0 & 0 & 1 \\
    0 & 0 & 1 & 0
  \end{array}
\right).
\end{eqnarray}
\end{document}